\documentclass{emulateapj}
\usepackage{xspace}
\newcommand{\minusone}{$^{-1}$\xspace}
\newcommand{\minustwo}{$^{-2}$\xspace}
\newcommand{\minusthree}{$^{-3}$\xspace}

\newcommand{\OIII}{[\textsc{Oiii}]\xspace}

\newcommand{\kms}{\ km s$^{-1}$\xspace}

\newcommand{\gal}{NGC~1068\xspace}
\newcommand{\HST}{\emph{HST}\xspace}
\newcommand{\fig}{Figure~}

\newcommand{\tab}{Table~}

\newcommand{\eqn}{Equation~}
\newcommand{\eqns}{Equations~}
\newcommand{\point}{$ \!\!.  $\thinspace}

\newcommand{\water}{H$_2$0\xspace}
\newcommand{\solarmass}{\mbox{\ M\ensuremath{_{\odot}}}\xspace}
\submitted{Accepted for publication in the Astrophysical Journal}

\shortauthors{Das et al.}  \shorttitle{Dynamics of the NLR in
NGC~1068}

\begin{document}

\title{Dynamics of the Narrow-Line Region in the Seyfert
  2 Galaxy \gal\altaffilmark{1}}

\author{V. Das\altaffilmark{2}, D.M. Crenshaw\altaffilmark{3},
  S.B. Kraemer\altaffilmark{4}}

\altaffiltext{1}{Based on observations made with the NASA/ESA Hubble
  Space Telescope. STScI is operated by the Association of Universities
  for Research in Astronomy, Inc., under NASA contract NAS5-26555.}

\altaffiltext{2}{Indian Institute of Astrophysics, 2$^{nd}$ Block,
  Koramangala, Bangalore-560034, alvin@iiap.res.in}

\altaffiltext{3}{Department of Physics and Astronomy, Georgia State
  University, Astronomy Offices, One Park Place South SE, Suite 700,
  Atlanta, GA 30303, crenshaw@chara.gsu.edu}

\altaffiltext{4}{Department of Physics, Catholic University of
  America, 200 Hannan Hall, Washington DC, MD 20064,
  stiskraemer@yancey.gsfc.nasa.gov.}

\begin{abstract}
  We present dynamical models based on a study of high-resolution
  long-slit spectra of the narrow-line region (NLR) in NGC 1068
  obtained with the Space Telescope Imaging Spectrograph (STIS) aboard
  \emph{The Hubble Space Telescope (HST)}. The dynamical models
  consider the radiative force due to the active galactic nucleus
  (AGN), gravitational forces from the supermassive black hole (SMBH),
  nuclear stellar cluster, and galactic bulge, and a drag force due to
  the NLR clouds interacting with a hot ambient medium. The derived
  velocity profile of the NLR gas is compared to that obtained from
  our previous kinematic models of the NLR using a simple biconical
  geometry for the outflowing NLR clouds. The results show that the
  acceleration profile due to radiative line driving is too steep to
  fit the data and that gravitational forces along cannot slow the
  clouds down, but with drag forces included, the clouds can slow down
  to the systemic velocity over the range 100--400 pc, as observed.
  However, we are not able to match the gradual acceleration of the
  NLR clouds from $\sim$ 0 to $\sim$ 100 pc, indicating the need for
  additional dynamical studies.
\end{abstract}

 ~~~~~

\section{Introduction}
Recent studies of the kinematics of the NLRs of Seyfert galaxies have
taken advantage of the high resolution of \HST to map the velocities
of these regions. \citep[and references
therein]{Evans1993,Macchetto,Hutchings1998,Nelson,Crenshaw4151,Crenshaw1068,Ruiz2001,Cecil,Ruiz2005,Das,Das2006}.
Although there have been a number of papers on the dynamical aspects
of the NLR, most of these studies relied on ground-based data limited
to spatial resolutions of $\geq$ 50 pc for even the most nearby AGN
\citep[see][and references therein]{Kaiser}. These studies relied
primarily upon spatially integrated line profiles to understand the
dynamics of the NLR \citep{Schulz,Veilleux1991}, but the problem arose
that the emission-line profiles of the NLR can be explained by many
different types of dynamical models, such as infall, rotation and
outflows
\citep{Capriotti1980,Capriotti1981,Vrtilek1983,Krolik1984,Vrtilek1985}.

With the launch of \HST and its high angular resolution ($\sim$
0\arcsec\point1), the NLRs of Seyfert galaxies have received
considerable attention. With the limited long-slit capability of the
faint object camera (FOC), and later the expanded capability of STIS,
detailed constraints on the kinematics of the NLRs in Seyfert and
other galaxies became possible. In turn, these kinematic studies
provided good diagnostics upon which dynamical analyses can be based.

\subsection{Previous Kinematics Studies of the NLR of Seyfert Galaxies}
The structure of the NLR resembles a bicone as expected from a simple
unified model of Seyfert galaxies, due to collimation by a thick torus
\citep{Antonucci}. Both \citet{Schulz} and \citet{Evans1993} have
modeled the NLR of NGC~4151 and found it to be consistent with a
biconical geometry. In an \HST study done on a sample of Seyfert 1 and
2 galaxies, \citet{Schmitt} compared both Seyfert types to study their
NLR morphologies, and found triangular structures in most of their
Seyfert 2s, and circular structures in most of their Seyfert 1s,
consistent with the unified model and biconical structure for the NLR
\citep[see also][]{Schmitt2003a,Schmitt2003b}. \citet{Veilleux2001}
modeled the inner regions of the Seyfert galaxy NGC 2992, and found
that the ionized gas can be fitted with a biconical structure.

We have recently completed a study of the kinematics of the NLRs in
two Seyfert galaxies (\citeauthor{Das} 2005, hereafter Paper I, and
\citeauthor{Das2006} 2006, hereafter Paper II). Since our dynamical
work in this paper is a direct extension of the works of those two
papers, we summarize our results below. In Paper I, kinematic models
were developed to match the emission-line velocities from
high-resolution STIS spectra within $\sim$ 400 pc of the central black
hole of the Seyfert 1 galaxy NGC~4151. The NLR gas showed strong
evidence of acceleration from $\sim$ 0 pc out to $\sim$ 100 pc, then
showed deceleration back to systemic velocity at $\sim$ 400 pc with
velocity roughly proportional to distance in each case. The maximum
velocity of the outflowing gas at the turnover point (96 pc) was
$\sim$ 800\kms relative to the black hole. Based on our kinematic
model, the NLR could be represented by a bicone, with inner and outer
half opening angles of 15 and 33\arcdeg\ respectively, and inclination
of $\sim$ 45\arcdeg\ with respect to the plane of the sky, consistent
with previous kinematic work done on NGC~4151 with different slit
positions \citep{Crenshaw4151}. Some of the fainter NLR clouds showed
evidence of backflow at the point where the clouds turnover in their
velocities. The radio jet was found to have little effect on the
kinematics of the NLR clouds, however there was some evidence of
radial velocity splitting of the fainter NLR clouds near bright knots
in the radio jet. The brighter clouds were not accelerated by the jet.

In Paper II, we developed a similar model for the Seyfert 2 galaxy
NGC~1068 again with high-resolution spectra taken with the STIS aboard
\HST. With seven parallel slit positions covering the entire NLR, we
extracted radial velocity profiles and matched them with our newly
developed 3-D biconical models. Our kinematical models showed that the
NLR gas accelerated out to $\sim$ 140 pc (the turnover point), then
decelerated back to systemic velocity at a distance of $\sim$ 400 pc
from the central black hole. The maximum velocity of outflow of the
gas was $\sim$ 2000\kms, with respect to the black hole, and the model
predicted and inner and outer half opening angles of the bicone of 20
and 40\arcdeg\ respectively, and an inclination of 5\arcdeg\ out of
the plane of the sky. We used high resolution radio maps of the NLR
obtained from \citet{Gallimore2004} to search for jet/cloud
interactions. Evidence showed that, similar to NGC~4151, the fainter
NLR clouds were split near bright knots in the radio jets, whereas the
brighter NLR clouds remained unaffected by the jet.

Other Seyfert galaxies show similar flow patterns to NGC~4151 and
NGC~1068, such as the Seyfert 2 galaxy Mrk 3
\citep{Ruiz2001,Ruiz2005}. The radial velocities of these galaxies
were matched with a common kinematic model with little variation in
the parameter space, which begs the question, what are the physical
processes involve that would cause such a similar flow pattern in both
types of Seyfert galaxies?  It is this question that motivated us to
carry out this dynamical study.

\subsection{Previous Dynamic Studies of the NLR of Seyfert Galaxies}
\label{flow}
In the first truly dynamical model of the NLR based on \HST
kinematics, \citet{Everett2006} attempted to fit the NLR velocities in
NGC~4151, based on measurements done by \citet{Das}. They tested an
isothermal Parker wind model which assumes thermally expanding winds
\citep{Parker1965}. By assuming a spherical cloud geometry, they let
the Parker wind drag along the embedded NLR clouds. As the Parker wind
accelerates by thermal expansion and slowly loses heat by adiabatic
cooling, the clouds are also accelerated to high speeds. They then let
the Parker wind run into a low density ambient medium to slow the
velocity of the wind, and hence the clouds, to the systemic velocity.
The model explains the velocity profile of the NLR of NGC~4151, but
suffers from the fact that an isothermal wind cannot be sustained out
to large distances. Also, the mass profile of the SMBH plus galaxy,
which determines the temperature profile for their model, is not
exactly known for NGC~4151. Hence while their model was not successful
on physical grounds, it is worth noting its relative success in
matching the NLR kinematics.

In this paper, the main question we want address is: how can we
constrain the dynamics of the NLR in Seyfert galaxies with the
detailed knowledge that we have gained from our kinematic studies?
Here, we concentrate on the dynamics of the NLR in NGC~1068, since it
has the best constraints. We will start with a simple construction of
the enclosed mass function based on data from previous studies and
eventually formulate a radiation pressure-gravity tug-of-war on the
NLR clouds. The questions we will attempt to answer include the
following: 1) If the NLR gas is in outflow, is radiation pressure
really the best driving mechanism? 2) If the NLR gas is turning over
its velocity and decelerating back to systemic, is gravity responsible
for stopping the gas? 3) Can we fit the velocity profile of the data
with a simple radiation-gravity law, or do we need to include another
force (such as drag)?

Our analysis applies to NGC~1068 in particular, but has relevance to
all Seyferts in general that show signs of gas outflow and subsequent
deceleration \citep{Ruiz2005}. To test whether gravity is playing any
role in stopping and turning back the gas velocity, we will construct
a mass profile within $\sim $ 10,000 pc from the nucleus of NGC~1068.
With the mass profile in hand, we will test whether the gas kinematics
are dominated by rotation. Such a test might prove fruitful for
rotation in NGC~4151, which shows redshifts northeast and blueshifts
southwest of the nucleus, but from a geometrical point of view,
rotation of the gas will prove difficult to match the velocity field
of NGC~1068, which shows blueshifts and redshifts on each side of the
nucleus. Next we plot the escape velocity with distance to see if the
gas should escape or not, given the velocities seen in the data, and
whether or not the kinematics of the NLR can be dominated by
gravitational infall. Then we concentrate on outflow assuming
spherical symmetry and pure radial motion. First we determine if the
deceleration of the gas can be attributed to the enclosed mass,
regardless of the outward accelerating force. We then apply radiative
line driving plus gravitational forces and compare the results to the
observed velocity law of NGC~1068 derived from the kinematic models.
Finally, we introduce a drag force due to an ambient medium on the NLR
clouds, in addition to radiation pressure and gravity, to determine if
it can improve the fit to either the accelerating or decelerating
portions of the observed velocity curve.

\section{Building the Mass Profile}
In building a model of the enclosed mass as a function of distance
from the central SMBH, we incorporate various subsystems into
our mass profile. These include contributions from the SMBH, the
nuclear stellar cluster, and the bulge. For all these systems, we
assume spherical symmetry for simplicity. We have assumed that the
stellar cluster and bulge extend all the way inward to the SMBH, which
may overestimate the mass close in. The size of the stellar cluster
was estimated to be $\sim$ 140 pc based on a study by
\citet{Thatte1997} and the bulge was assumed to extend up to 10000 pc.
Mass contribution from the galactic disk of NGC~1068 was neglected
because the galactic potential is dominated by the large bulge to at
least 1000 pc, which is well beyond the extent of the NLR.

\subsection{The Black Hole Mass}
NGC~1068 is one of only a few AGN that shows an edge-on disk of \water
maser emission close to the SMBH. The disk shows the signature of a
rotational velocity curve, which can be used to determine the mass of
the SMBH. \citet{Greenhill1997} estimated the mass of NGC~1068 to be
1.5$\times$10$^7$\solarmass within 0.65 pc, based on the velocity
field of the \water maser emission observed with the VLBA and the VLA.
We will therefore use this estimate for the mass of the SMBH.

\subsection{The Bulge Mass}
Since we are going out to $\sim $ 10,000 pc, which encloses the NLR
and the ENLR clouds, we need an accurate assessment of the total mass
within this region. At small distances ($\leq$ 1 pc), the SMBH is
dominant, although in the case of NGC~1068, a concentrated stellar
cluster is also providing substantial gravity close in. For an
estimate of the bulge mass in NGC~1068, we rely on the work of
\citet{Haring2004}. They found a tight correlation between black-hole
mass and bulge mass for a sample of 30 galaxies, including NGC~1068.
They determined the bulge mass by modeling the bulge with the Jeans
equation in spherical form. They assumed the bulge to be isotropic and
spherically symmetric, which might lead to an overestimation of the
bulge mass; however they also neglect any contribution from dark
matter, which would tend to underestimate the bulge mass. Their value
for the bulge mass in NGC~1068 is 2.3$\times$10$^{10}$\solarmass
within a radius $r$ = 3R$_e$ (3 effective radii), where R$_e$ = 3.1
$\pm$ 0.8 kpc, a value taken from the surface brightness deconvolution
of \citet{Marconi2003}. The effective radius R$_e$ is defined to be
such that half of the total light from the galaxy is predicted to be
contained within the isophotal ellipse that has area $\pi
\mathrm{R}_e^2$ \citep{Binney1998}.

Elliptical galaxies' and bulges' surface brightnesses can be well
described by the empirical formula developed by
\citet{deVaucouleurs1948}
\begin{equation}
  \label{sbp}
  I(R)=I_ee^{-7.6692[(\frac{R}{R_e})^\frac{1}{4}-1]},
\end{equation}
where $I_e$ = I($R_e$). In the 1980s--1990s, a family of stellar
density curves emerged that modeled both elliptical galaxies and
bulges well. These curves are of the form
\begin{equation}
  \label{dp}
  \rho(r)=\frac{(3-\gamma)M}{4\pi}\frac{\eta}{r^\gamma(r+\eta)^{4-\gamma}},
\end{equation}
where $\eta$ (in pc) is a scaling radius and M is the total mass of
the bulge \citep{Dehnen1993}. The parameter $\gamma$ determines
different types of models, where the $\gamma$ = 2 cases corresponds to
previous density models by \citet{Jaffe1983} and the $\gamma$ = 1
cases to models by \citet{Hernquist1990}. These models, when
integrated over a spherical volume, yield the following enclosed mass
profile:
\begin{align*}
  \label{mp}
  M(r) &= \int_0^r4\pi t^2 \rho(t)dt =
  \frac{4\pi(3-\gamma)M\eta}{4\pi}\int\frac{t^2dt}{t^\gamma(t+\eta)^{4-\gamma}}\\
  &= M\left(\frac{r}{r+\eta}\right)^{3-\gamma}.\tag{3}
\end{align*}
\setcounter{equation}{3}
For the special case when $\gamma$ = 1.5, the density profile of
\eqn\ref{dp} yields a surface density distribution that closely
matches the de Vaucouleurs surface brightness profile of
\eqn\ref{sbp} to within 15\% over nearly 4 decades in radius
\citep{Dehnen1993}. Therefore we adopt the following form of the mass
function as the bulge profile
\begin{equation}
  \label{mf}
  M(r) = M\left(\frac{r}{r+\eta}\right)^{1.5}.
\end{equation}
We find a suitable value for $\eta$ in \eqn\ref{mf} by first defining
the `half-mass radius' $r_{\frac{1}{2}}$:
\begin{equation}
  \frac{1}{2}M = M\left(\frac{r_\frac{1}{2}}{r_\frac{1}{2}+\eta}\right)^{\frac{3}{2}},
\end{equation}
which yields the following relation for $\eta$
\begin{equation}
  \label{cona}
  \eta = r_\frac{1}{2}(2^{\frac{2}{3}}-1).
\end{equation}
\citet{Dehnen1993} has found a simple approximation for
$\frac{R_e}{r_\frac{1}{2}}$ that depends only slightly on $\gamma$.
For a $\gamma \leq \frac{5}{2}$, he found that
\begin{equation}
  \frac{R_e}{r_\frac{1}{2}} \approx
  (0.7549-0.00439\gamma+0.00322\gamma^2-0.00182\gamma^3).
\end{equation}
Therefore we can find a value for $\eta$ by using $\gamma$ = 1.5,
$R_e$ given above, and \eqn\ref{cona}. We find that $\eta \sim 2400$
pc. We also know that M(3R$_e$) = 2.3$\times$10$^{10}$\solarmass, so
that
\begin{align*}
  M &=
  2.3\times10^{10}/\left[\frac{3\times3.1\times10^3}{3\times3.1\times10^3+2400}\right]^{1.5}\\
  &= 3.2\times10^{10}\solarmass.\tag{8}
\end{align*}
\setcounter{equation}{8}
The bulge mass distribution can finally be written as
\begin{equation}
  M(r) = 3.2\times10^{10}\left(\frac{r_{pc}}{r_{pc}+2400}\right)^{1.5}\solarmass.
\end{equation}

\subsection{The Nuclear Stellar Cluster Mass}
It is known that NGC~1068 has a compact nuclear stellar cluster $\sim
$ 140 pc in radius \citep{Thatte1997,Crenshaw1068a}, which contributes
significant mass to the total mass profile of the NLR of NGC~1068.
Therefore this mass must be taken into account when deriving the mass
profile. \citet{Thatte1997} found a mass of
6.8$\times$10$^8$\solarmass within 1\arcsec\ of the SMBH of NGC~1068,
assuming a virialized, isotropic, and isothermal distribution of the
stars. They used the stellar velocity distribution ($\sigma_*$) found
in \citet{Dressler1984}, a value of 143 $\pm$ 5\kms at $\sim $
1\arcsec, to calculate the total dynamical mass within 1\arcsec\ of
the nucleus of NGC~1068.
\begin{equation}
  M_{dyn} = \frac{2\sigma_*^2R}{G},
\end{equation}
where $R = 1\arcsec \approx 72$ pc for NGC~1068. This mass includes
contributions from the stellar cluster, the nucleus, and the bulge
within a radius of 72 pc. Therefore to find just the mass from the
stellar cluster $M_{sc}$, we took out the rest of the mass
contribution from the total mass.
\begin{equation}
  M_{sc}(72 pc) = M_{dyn}(72 pc) - M_{smbh} - M_{bulge}(72 pc)
\end{equation}
Making the various substitutions we have
\begin{align*}
  M_{sc}(72 pc) &= 6.8\times10^8 -
  1.5\times10^7 \\&-
  3.2\times10^{10}\left(\frac{72}{72+2400}\right)^{1.5} \\&=
  5.1\times10^8\solarmass.\tag{12}
\end{align*}
\setcounter{equation}{12}

Based on the radial surface brightness profile presented in \fig 4 of
\citet{Thatte1997}, \citet{beckert2004} computed a power law
consistent with the form $S_*(r) \propto r^{-1}$ to fit the data. They
claim that if the profile traces the stellar mass distribution, then
they would expect that the mass profile would have the form M$_*(r)
\propto r$, given a spherical, isothermal distribution of stars in
the cluster. Using their distribution we can estimate the stellar mass
function based on the condition $M_{sc,\,72} =
5.1\times10^8$\solarmass = $kr$, so that $k = \frac{5.1\times10^8}{72}
= 7.1\times10^6$\solarmass\,pc\minusone. The stellar cluster mass
function is therefore
\begin{equation}
  M_{sc}(r) = 7.1\times10^6r_{pc}\solarmass.
\end{equation}
Beyond the observed extent of the stellar cluster, we assumed that
$M_{sc}(r)$ is a constant. Finally the total enclosed mass function
for the NLR of NGC~1068 is given by $M_{tot}(r) =
M_{smbh}+M_{bulge}+M_{sc}$ or
\begin{align*}
  M_{tot}(r) &= 1.5\times10^7 +
  7.1\times10^6r_{pc}\\
  &+ 3.2\times10^{10}\left(\frac{r_{pc}}{r_{pc}+2400}\right)^{1.5}\solarmass.\tag{14}
\end{align*}
\setcounter{equation}{14}

\section{Dynamics Based on Gravity}
\begin{figure}[!ht]
  \vspace*{7mm}
  \hspace*{-5mm}
  \centering \includegraphics[width=0.43\textwidth,height=0.82\textheight]
  {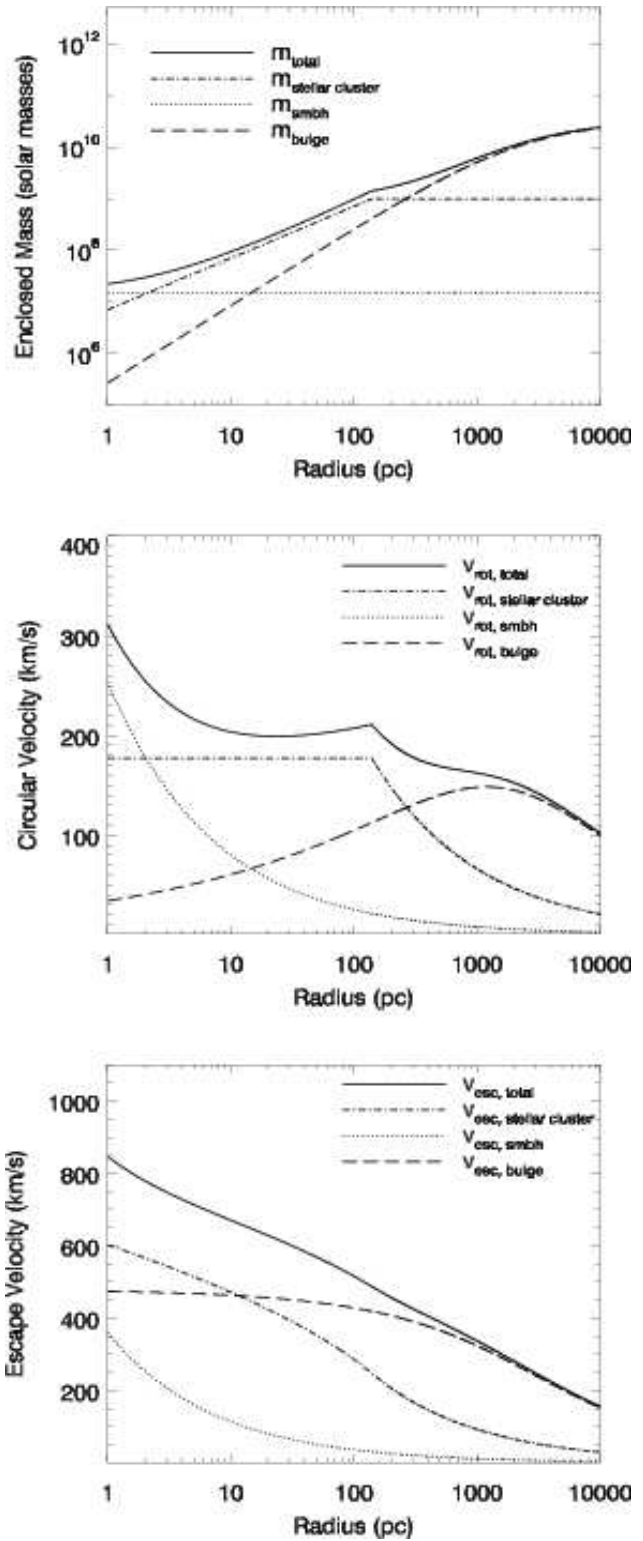}
  \caption{Top: The total enclosed mass profile of NGC~1068
    in solid black. The individual contributions to the total mass is
    also shown in this plot. They are the stellar cluster (dash-dot),
    the bulge (dash), and the SMBH (dot). The stellar
    cluster was modeled to have an extent of 140 pc.\\ Middle: The total
    rotational velocity profile of the NLR of NGC~1068 in solid black
    color. The individual rotation components due to the various masses
    are also shown.\\ Bottom: The escape velocity as a function of
    distance for the NLR of NGC~1068.}
  \label{mass_rot_esc_vel_pro}
\end{figure}

A figure representing the total mass enclosed within $r_{pc}$ is shown
in the top panel of \fig\ref{mass_rot_esc_vel_pro}.
The mass profiles for each contribution, the SMBH, bulge, and cluster
are also shown in the figure. Close in toward the nucleus, the nuclear
stellar cluster dominates up to its entire extent, while the bulge
takes over from there. The black hole mass dominates at $\leq$ 2 pc.
The kink in the total mass curve at $\sim $ 140 pc is because the
stellar cluster was cut off abruptly at this location. We could have
modeled the cluster to assume an exponential drop-off after 140 pc,
but this would not have contributed much to the gravitational force
exerted by the total mass.

\subsection{Rotation}
With the total mass profile we calculate the circular rotational
velocity and plot it as a function of distance as shown in the middle
panel of \fig\ref{mass_rot_esc_vel_pro}. The rotation velocity only
depends on the enclosed mass at radius $r_{pc}$, and is given by the
formula below
\begin{equation}
  V(r) = \sqrt{G\frac{M(r)}{r}}.
\end{equation}
For demonstration purposes, we show the rotational velocities as if
only each mass component was present, as well as the velocity for the
total mass profile. The rotational velocity profile indicates to us
that rotation cannot dominate the kinematics of the NLR of NGC~1068
because many observed data points exhibit large velocities ($\geq$
1000\kms), whereas the rotation curve never exceeds $\sim$ 220\kms at
large distances (100 pc). Again the stellar cluster dominates the
velocities up to $\sim $ 300 pc.

\subsection{Escape and Infall Velocity}
The escape velocity at a given distance $r_{pc}$ is calculated
numerically from the formula
\begin{equation}
  V(r) = \sqrt{\int_{r}^\infty 2G\frac{M(t)}{t^2}dt}\ \textrm{km\,s}^{-1},
\end{equation}
based on the enclosed mass function, and is plotted in the bottom
panel of \fig\ref{mass_rot_esc_vel_pro}. According to our kinematic
model, the maximum velocity at the turnover radius $r_t$ = 140 pc is
2000\kms (Paper II); therefore \fig\ref{mass_rot_esc_vel_pro}
tells us that the NLR clouds should have escaped after 140 pc, where
the escape velocity is only $\sim $ 500\kms. In the data however,
clouds at 140 pc are at much higher velocity than escape velocity; yet
after the turnover point, the clouds start to decrease their velocity
and return to systemic. Thus, some force other than gravity is causing
the clouds to decelerate at r $\geq$ 140 pc. The infall velocity
profile is equivalent to the escape velocity profile, except that the
velocity vector is now directed inward. Therefore gravitational infall
also cannot account for the faster moving clouds at 140 pc, and in
general does not match the observed velocity profile (Paper II).

\subsection{Gravitational Drag}
\begin{figure}[!ht]
  \vspace*{7mm}
  \centering \includegraphics[width=0.45\textwidth,height=0.57\textheight]
  {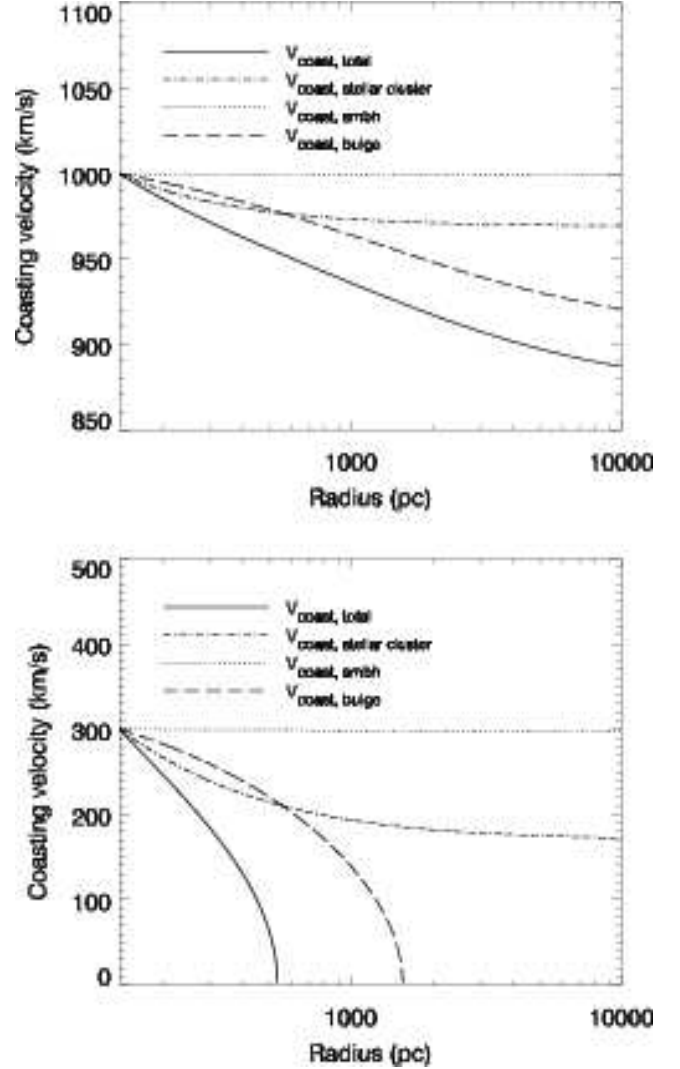}
\caption{The velocity slowing down with distance for an
  initial launch at 140 pc with 1000\kms (top) and 400\kms (bottom)
  shown in solid lines for the clouds in NGC~1068. The other mass
  components' effect are shown in different linestyles represented by
  the key above the curves. The clouds barely slow down out to 10,000
  pc when launch with 1000\kms. However, gravity was able to slow down
  the clouds down when they were launched with 300\kms.\vspace*{2mm}}
  \label{coasting_velocity_profiles}
\end{figure}

To test the importance of the force of gravity alone on slowing down
the outflowing NLR clouds, we give the clouds a maximum velocity at
the turnover point and let gravity do the rest. In other words, we
assume that there is no outward driving force after the turnover point
and let the clouds coast under the force of gravity. The top panel of
\fig\ref{coasting_velocity_profiles}
shows that with a maximum velocity $\geq$ 1000\kms at 140 pc, there is
little deceleration with radius. However, with maximum velocities
$\leq$ 300\kms gravity can slow the clouds down to rest, as seen in
the bottom panel of \fig\ref{coasting_velocity_profiles}. The maximum
velocities in the NLRs of some Seyfert galaxies are on the order of
$\sim $ 400\kms \citep{Ruiz2005}, and gravitational deceleration may
be important in these cases. The kinematic model of NGC~1068, however,
shows maximum velocities of up to $\sim $ 2000\kms, clearly out of the
reach of gravitational deceleration. Gravity alone cannot slow down
the outflowing clouds in this case and there must be some other force
or forces involved.

To compound the problem, suppose we let radiation or some other force
push on the gas while gravity is trying to pull it back. In this case
the gravitational deceleration will be even less. However, to further
explore this issue, we generate a line-driven radiation model, with
gravity competing to slow the gas down. This will eventually lead us
in the direction of including a drag force to complete the analysis.

\section{Radiation-Gravity Formalism}
The various mechanisms to push the gas out from close to the nucleus
include a radiation pressure driven wind, a thermally driven wind, or
a magneto-hydrodynamic (MHD) wind \citep{Crenshaw2003}. The latter two
methods are discussed in more detail in \citet{Everett} and
\citet{Everett2006}, and were summarized in \S\ref{flow}. However, no
dynamical model to date has led to a satisfactory description of the
kinematics in NGC~1068 or NGC~4151. The radiation driven wind
mechanism is most efficient when the momentum imparted to the gas is
due to line-driving (bound-bound transitions), although bound-free and
free-free electron transitions (Thomson scattering) also contribute to
driving the gas out \citep{Chelouche2001}. The radiation force is
dependent on the ionization state of the gas, with lower ionization
states more efficient due to the greater availability of electrons in
the bound states. If dust is mixed in with the NLR gas, then it will
compete with the gas in absorbing ionizing photons and hence radiation
pressure on the dust can become an important contributor to the
velocities of the outflowing gas in the NLR \citep{Dopita2002}. If the
dust grains are electrically charged, they can drag the ionized gas
along to similar velocities as the dust. In this section we ignore the
effects of dust, which would only increase the radiative acceleration.
Thus we consider a radiation driving mechanism coupled with the
effects of gravity to find the velocity profile of the NLR gas.

\subsection{Building the Velocity Equation} We start with the
acceleration due to radiation on a point mass,
\begin{equation}
  \label{acc}
  a(r) = \frac{L\sigma_T\mathcal{M}}{4\pi r^2cm_p},
\end{equation}
where $a$ is the acceleration, $L$ is the bolometric luminosity of
NGC~1068, $\sigma_T$ is the Thomson scattering cross section for the
electron, $r$ is the distance, $c$ is the speed of light, $m_p$ is the
mass of the proton, and $\mathcal{M}$ is the force multiplier. As
mentioned above, to really drive the gas out efficiently, we need to
incorporate other sources of opacity such as bound-bound and
bound-free opacity in addition to those from Thomson scattering. These
additional opacities are included via the force multiplier,
$\mathcal{M}$ in \eqn\ref{acc}, which is primarily a function of
ionization parameter\footnote{The ionization parameter $U$ is defined
  as $U = \frac{\int_{\nu_0}^{\infty}\frac{L_{\nu}}{h\nu}d\nu}{4\pi
    r^2n_Hc}$ = $\frac{\textrm{\# of ionizing photons}}{\textrm{\# H
      atoms}}$ at the ionized face of the clouds, where $h\nu_0$ =
  13.6 eV and $n_H$ is the hydrogen number density.} $U$ for a given
spectral energy distribution \citep[][and references
therein]{Crenshaw2003}.

The acceleration due to gravity per mass is simply given by
\begin{equation}
  \label{acc2}
  a(r) = -\frac{GM(r)}{r^2},
\end{equation}
where $M(r)$ is the total enclosed mass within $r$ parsecs, and $G$ is
the universal gravitational constant. Putting \eqns\ref{acc} and
\ref{acc2} together we have
\begin{equation}
  \label{acc3}
  a(r) = \frac{L\sigma_T\mathcal{M}}{4\pi r^2cm_p} -\frac{GM(r)}{r^2}.
\end{equation}
Now we will have to rewrite the acceleration in terms of velocity as a
function of radius, and then solve for $v(r)$.
\begin{equation}
  a = \frac{dv}{dt} =
  \frac{dr}{dt}\frac{dv}{dr} = v\frac{dv}{dr}.
\end{equation}
Therefore we can now write \eqn\ref{acc3} as a simple separable
differential equation
\begin{equation}
  \label{vel2}
  vdv = \frac{L\sigma_T\mathcal{M}}{4\pi r^2cm_p}dr -
  \frac{GM(r)}{r^2}dr.
\end{equation}
Substituting for the constants in \eqn\ref{vel2} and converting to
appropriate units of \kms and pc, then integrating and setting the
initial velocity to zero yields the following form for $v(r)$
\begin{equation}
\label{finvel}
  v(r) = \sqrt{\int^{r}_{r_1}\left[6840L_{44}\frac{\mathcal{M}}{t^2} -
    8.6\times10^{-3}\frac{M(t)}{t^2}\right]dt},
\end{equation}
where $L_{44}$ is luminosity in units of $10^{44}$ ergs s$^{-1}$ and
$M(r)$ is in units of $M_{\odot}$. The constraints on the luminosity
and the force multiplier are presented in the next section.

\subsection{Physical Constraints on the NLR of NGC~1068}
Since NGC~1068 is a Seyfert 2 galaxy, we cannot measure its luminosity
directly. From \citet{Pier1994}, the total luminosity of NGC~1068 is
given by
\begin{equation}
  \label{lum}
  L_{bol} =
  2.2\times10^{11}\left(\frac{f_{refl}}{0.01}\right)^{-1}\left(\frac{D}{22Mpc}\right)^2L_{\odot},
\end{equation}
where $f_{refl}$ is the fraction of nuclear flux observed as scattered
radiation, $D$ is the distance to NGC~1068, and $L_{\odot}$ is the
solar luminosity. The most uncertain term in \eqn\ref{lum} is
$f_{refl}$. \citet{Pier1994} have summarized a range of values for
$f_{refl}$ that have been determined previously by several authors.
The range in $f_{refl}$ spans a few orders of magnitude, from
0.001--0.05. They claim that the best estimate comes from
\citet{Miller1991}, who had determined a value for $f_{refl}$ to be
0.015, based on observations of \OIII and broad $H\beta$ luminosity
and their ratio. \citet{Pier1994} concluded that $f_{refl}$ is
probably within a factor of a few of 0.01, hence we have adopted a
value for $f_{refl}$ of 0.015 because it is the ``best'' estimate and
close to the average value adopted by \citet{Pier1994}. We already
know the distance to NGC~1068 as 14.4 Mpc \citep{Bland-Hawthorne} so
the bolometric luminosity of NGC~1068 is given by
\begin{align*}
  L_{bol} &=
  2.2\times10^{11}\left(\frac{0.015}{0.01}\right)^{-1}\left(\frac{14.4}{22Mpc}\right)^2
  L_{\odot}\\ &= 2.4\times10^{44}\ \textrm{erg\,s}^{-1},\tag{24}
\end{align*}
a value which could be uncertain by a factor of 0.3--15, depending on
$f_{refl}$.
\setcounter{equation}{24}

The emission lines arising from the NLR gas are best fitted with a two
component photoionization model at each position, based on \HST/STIS
long-slit spectra \citep{Kraemer2000c}. According to
\citeauthor{Kraemer2000c}, the ionization state of the gas ranges from
$U \sim\ 10^{-1.5}$--$10^{-3.0}$ for the two components but seems to
vary little with distance. Using $U$ and the SED for NGC~1068, we
found the force multiplier to vary from $\mathcal{M} \approx $
500--6000 for the front face (ionized face) of the clouds, based on
CLOUDY models \citep{Ferland}. The mass function was derived in
previous sections, and $L_{44}$ = 2.4, so we can now numerically solve
\eqn\ref{finvel} for $v(r)$, assuming $\mathcal{M}$ is constant with
distance. The results are presented in the next section.

\subsection{Radiation and Gravity Results}
\begin{figure}[!ht]
  \vspace*{7mm}
  \hspace*{-7mm}
  \centering \includegraphics[width=0.45\textwidth,height=0.79\textheight]
  {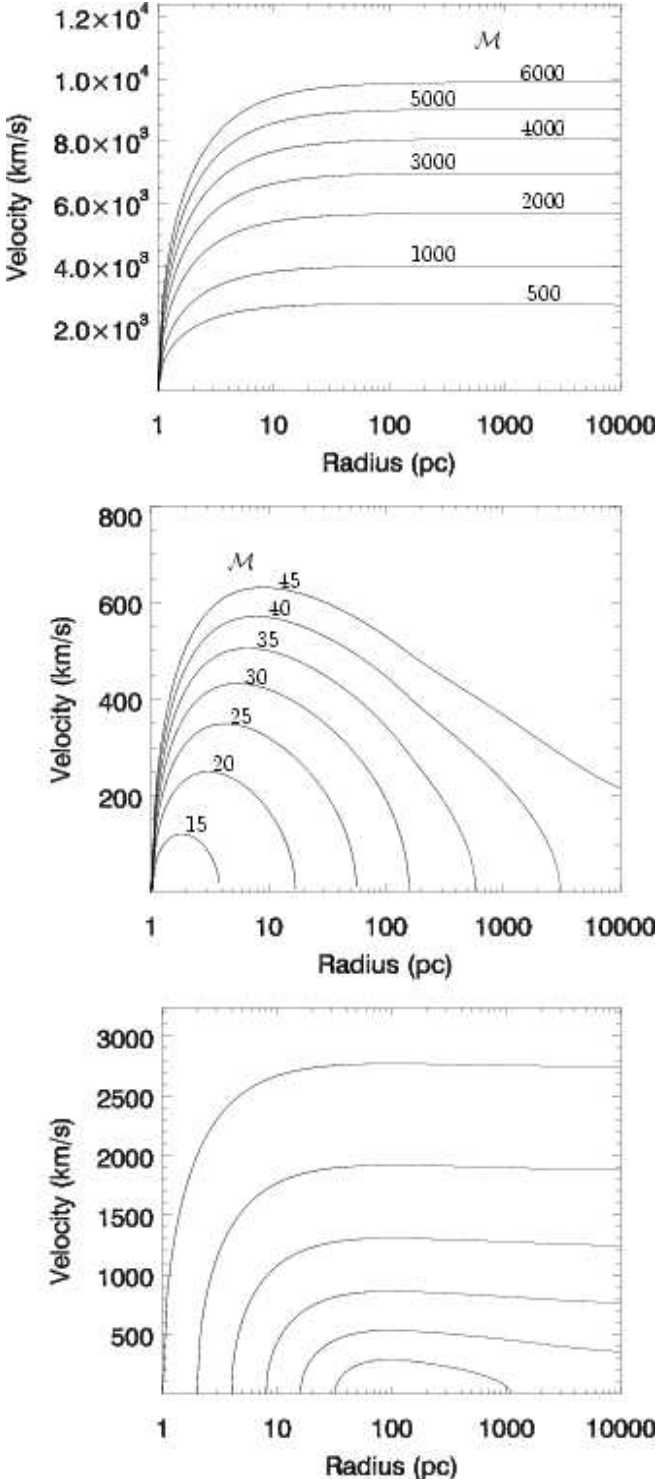}
\caption{Top: The velocity profiles for several force
  multipliers $\mathcal{M}$ with a fixed launch radius $r_1$ = 1 pc.
  None of the curves show a significant decrease in velocity up to
  10,000 pc.\\ Middle: The velocity profiles for several lower force
  multipliers with $r_1$ = 1 pc. Curves with force multipliers less
  than 40 turn over before reaching 10,000 pc, although their maximum
  velocities do not fit the data of NGC~1068.\\ Bottom: The velocity
  profiles for several different launch radii (top to bottom curves
  r$_1$ = 1, 2, 4, 8, 16, 32 pc) all modeled with a constant force
  multiplier of 500. The bottom curve was able to turn over and return
  to systemic at $\sim$ 1000 pc. The maximum velocity and maximum
  distance reached for these curves do not fit the data.}
  \label{fm_high_low_Ro_velocity_profiles}
\end{figure}

\eqn\ref{finvel} has only two parameters that we can vary to find the
velocity $v$ as a function of distance $r$: the launch radius $r_1$,
and the force multiplier $\mathcal{M}$. We plotted $v(r)$ for various
combinations of launch radii and force multipliers of the gas. The top
panel of \fig\ref{fm_high_low_Ro_velocity_profiles}
shows that with a launch radius of $r_1$ = 1 pc, the velocity of the
gas increases with force multiplier, but quickly reaches a terminal
velocity and does not slow down significantly, even out to $\sim $
10,000 pc. This is not a surprise, as previous plots had shown that
the mass is not enough to slow down the high-velocity clouds. In
addition, the maximum velocities for $\mathcal{M} \geq$ 500 are too
high compared to the observations, indicating that we must increase
the launch radius $r_1$.

The physical constraints on the NLR gas suggested that the force
multiplier is larger than 500. However, we generated plots for
$\mathcal{M}$ smaller than 500 to see whether or not gravity can slow
the clouds down, if the clouds are launched from 1 pc. In the middle
panel of \fig\ref{fm_high_low_Ro_velocity_profiles}, we see that with
force multipliers $\leq$ 40, gravity does slow the clouds down, but
that the maximum velocities are too low to fit the data of NGC~1068.
Such models may prove useful for Seyferts with lower outflow
velocities in their NLRs, such as NGC~4151 and Mkn~3
\citep{Das,Ruiz2001}.

We next fix the force multiplier to the lowest value of 500 (because
higher values of $\mathcal{M}$ will be even more problematic to slow
down the clouds) and vary the launch radius. These plots are presented
in the bottom panel of \fig\ref{fm_high_low_Ro_velocity_profiles}. The
maximum velocity falls below the observed values at $r_1 \geq$ 2 pc,
and the final velocity does not return to zero until we reach large
launch radii. As the launch radius increases to $\sim $ 32 pc, the
maximum attainable velocity of the outflowing clouds decreases, and
the enclosed mass is finally able to turn the velocity around. However
by that time, the maximum velocity is much lower than predicted by the
kinematic model. Also, the launch radius and the maximum distance of
outflow become too large to fit the data, since the NLR clouds
generally launch from close in and drop in radial velocity within 400
pc (Paper II). Clearly these curves cannot fit the observed NLR
velocities of NGC~1068.

\section{Radiation, Gravity, and Drag Forces}
The radiation-gravity interaction on the NLR clouds of NGC~1068 fails
to reproduce its velocity profile. For reasonable parameters, the
velocity increases quickly close to the nucleus and mostly remains
constant over large distances regardless of launch radius. The maximum
outflow velocity is rather sensitive to launch radius and decreases
with increasing $r_1$ but the clouds' velocities never turn over and
decrease. Fine-tuning $r_1$ and $\mathcal{M}$ outside of the range of
reasonable parameters can lead to a deceleration profile, but the
resulting velocity profile and amplitude does not match the observed
trend. Therefore we conclude that there must be additional forces at
play in the NLR to account for the velocity profile that we see in the
data. One such force that could explain the trend in the data is drag,
whereby the clouds are slowing down in a more diffuse, hotter, and
higher ionization medium \citep{Crenshaw1068}.

\subsection{Deceleration due to Drag}
The drag force exerted on a cloud by an ambient medium is
\begin{equation}
  F_{drag,\,cloud} = \rho_{med}(v_{cloud} - v_{med})^2A_{cloud},
\end{equation}
where $\rho_{med}$ is the mass density of the ambient medium,
$v_{cloud}$ and $v_{med}$ are the velocities of the cloud and medium
respectively, and $A_{cloud}$ represents the cross sectional area of a
cloud \citep{Everett2006}. Following Everett \& Murray, we assume the
clouds to be spherical with mass $m_{cloud}$ = $\frac{4}{3}\pi
R^3_{cloud}\rho_{cloud}$, where $R_{cloud}$ is the radius of a cloud,
and $\rho_{cloud}$ is its mass density. The acceleration on the clouds
due to drag is then
\begin{align*}
  a_{drag,\,cloud} &= \frac{\rho_{med}(v_{cloud} -
    v_{med})^2\pi R^2_{cloud}}{\frac{4}{3}\pi R^3_{cloud}\rho_{cloud}}\\
    &= \frac{m_pn_{med}}{m_pn_{cloud}R_{cloud}}\frac{3}{4}\left(v_{cloud}-v_{med}\right)^2,\tag{26}
\end{align*}
\setcounter{equation}{26}
where $n$ is the hydrogen number density and $m_p$ is the mass of the
proton. In our case, we assume that the velocity of the ambient medium
is zero and the radius of the cloud remains constant. Therefore with
$v_{med} \approx 0$ we can write
\begin{equation}
  a_{drag,\,cloud} = -\frac{n_{med}}{N_{H,\,cloud}}\frac{3}{4}v^2_{cloud},
\end{equation}
where $N_{H,\,cloud}$ ($\approx n_{cloud}R_{cloud}$) is the hydrogen
column density of the cloud, and the minus sign represents
deceleration. Combined with the radiative and gravitational
acceleration of \eqn\ref{acc3}, the total acceleration on the clouds
becomes
\begin{equation}
  a_{tot} = \frac{L\sigma_T\mathcal{M}}{4\pi r^2cm_p} -\frac{GM(r)}{r^2} -
  \frac{n_{med}}{N_{H,cloud}}\frac{3}{4}v^2_{cloud},
\end{equation}
and in differential form is
\begin{equation}
  vdv = \frac{L\sigma_T\mathcal{M}}{4\pi r^2cm_p}dr - \frac{GM(r)}{r^2}dr -
  \frac{n_{med}}{N_{H,cloud}}\frac{3}{4}v^2dr.
\end{equation}
Substituting the various constants and converting to units of pc and
\kms, we have the following inseparable differential equation to solve
\begin{align*}
  \label{diff}
  \frac{dv(r)}{dr} &= 3420L_{44}\frac{\mathcal{M}}{v(r)r^2} -
  4.3\times10^{-3}\frac{M(r)}{v(r)r^2}\\ &-
  2.3\times10^{-2}\frac{n_{med}}{N_{20}}v(r),\tag{30}
\end{align*}
where $N_{20} = N_H/10^{20}\textrm{cm}^{-2}$.
\setcounter{equation}{30}

\subsection{Constraints on Cloud and Medium Densities for NGC~1068}
Both NGC~1068 and NGC~4151 show evidence for highly ionized gas
extended throughout their NLRs, based on Chandra X-ray Observatory
images \citep{Ogle,Ogle2003}. Thus, we assume an ambient medium that
is highly ionized, with ionization parameter $U \approx $ 10. The
ionization parameter, which is inversely related to the density and
radius, can be written as follows
\begin{equation}
U \propto \frac{\int_{\nu_0}^{\infty}\frac{L_{\nu}}{h\nu}d\nu}{r^2n_H}
\end{equation}
Therefore if the ambient and the NLR clouds see the same ionizing
luminosity (L$_{ion}$) at a particular distance $r$ from the source,
we can write
\begin{equation}
\label{ratio}
n_{med} = n_{cloud}\frac{U_{cloud}}{U_{med}}.
\end{equation}
\citet{Kraemer2000c} provided good constraints on the parameters on
the right side of \eqn\ref{ratio}. The densities of the NLR clouds are
almost constant out to large distances from the nucleus with a typical
value of $n_{cloud} \approx$ 10$^4$ cm\minusthree for clouds with
$U_{cloud} \approx$ 10$^{-3}$. If we substitute these estimates in
\eqn\ref{ratio}, we will have a typical estimate for $n_{med}$ as
follows
\begin{equation}
n_{med} = 10^4\frac{10^{-3}}{10} = 1\ \textrm{cm}^{-3}.
\end{equation}
\citeauthor{Kraemer2000c} found column densities for the NLR clouds in
the range $N_H$ = 10$^{19}$--10$^{21}$ cm\minustwo, which corresponds
to $N_{20}$ = 0.1--10. Since we are interested in the ratio
$\frac{n_{med}}{N_{20}}$, varying either parameter while keeping the
other constant will result in the same curves. In this paper, we
choose to keep $N_{20}$ constant at the average value of 1 and vary
$n_{med}$, because $n_{med}$ is the most unknown quantity. We already
know that $\mathcal{M}$ can take values from 500--6000, so
\eqn\ref{diff} can now be solved numerically for $v(r)$ with various
values of the launch radius ($r_1$), force multiplier ($\mathcal M$),
and the ratio $\frac{n_{med}}{N_{20}}$. We used Mathematica v5.2,
which employs the most efficient choices among various flavors of
Runge-Kutta algorithms, to solve for $v(r)$. The results are presented
in the next section.

\subsection{Radiation, Gravity and Drag Results}
\begin{figure}[!ht]
  \vspace*{7mm}
  \hspace*{-5mm}
  \centering
  \includegraphics[width=0.45\textwidth,height=0.55\textheight]
  {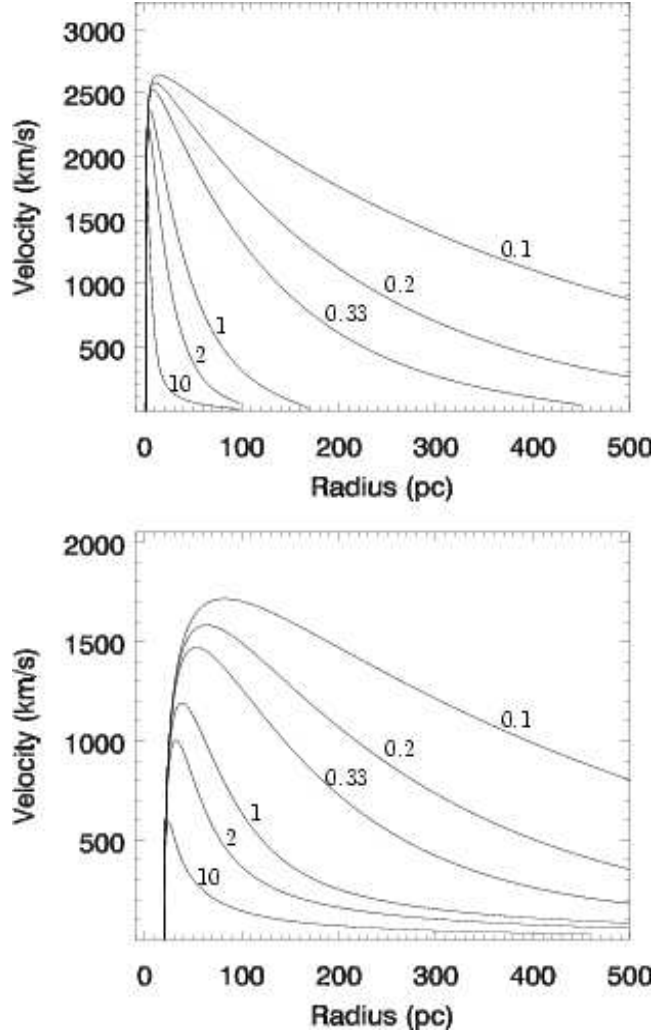}
  \caption{Top: The velocity profiles for force multiplier 500, a
  launch radius of 1 pc, and varying medium densities $n_{med}$ shown
    by the numbers.\\ Bottom: The velocity profiles for a force
    multiplier of 6000, a launch radius of 20 pc, and varying medium
    densities similar to the top panel.}
  \label{fm_500_r_1_fm_6000_r_20}
\end{figure}

The top panel of \fig\ref{fm_500_r_1_fm_6000_r_20}
represents several plots with a force multiplier of 500, a launch
radius of 1 pc, a column density of $N_{20}$ = 1, and various
densities of the ambient medium. The figure shows that when launched
at 1 pc, the gas accelerates to a maximum velocity inside 10 pc, then
slows down again.  The velocity slows down faster with increasing drag
forces, as measured by increasing densities of the ambient medium.
However the point of maximum velocity is too close in to match the
data, which has a turnover velocity at $\sim $ 140 pc. The curve with
medium density $n_{med}$ = 0.33 shows gradual deceleration out to
$\sim $ 500 pc, but its maximum velocity is slightly too high. Any
other curve has either too high velocity at turnover or the velocity
drops too quickly. Note that the velocity of the data we are trying to
match has $\sim $ 2000\kms at a turnover of $\sim $ 140 pc for
NGC~1068. With force multipliers higher than 500, the same trend as in
\fig\ref{fm_500_r_1_fm_6000_r_20} is seen in the velocity model,
except that the velocities are much higher ($\geq$ 6000\kms). Again,
the curve that seems best to match the deceleration part in the data
is with medium density $n_{med}$ = 0.333. The rest of the curves
decelerate too quickly or too slowly. The major problem with all of
these curves is that the velocity turnover point is much closer to the
nucleus than 140 pc.

Another way to decrease the overall velocity is to increase the launch
radius, and tweak the force multiplier and ambient density. We tried a
launch radius of 10 pc with force multiplier ranging from 500--6000.
The maximum velocity drops as expected, reaching $\sim$ 700\kms for a
force multiplier of 500 and climbs to $\sim$ 2500\kms when we increase
the force multiplier to 6000. We noticed that the turning point
increases by a factor of 5 as we increase the launch radius from 1 to
10 pc. However the turnover point is still too low to fit the data
well. In order to have a good fit, we first need to maximize the
launch radius and then tweak the force multiplier and medium density
to get the closest match to the data as possible.

The resolution in our data is $\sim $ 10 pc and we should not launch
much beyond this distance since we see clouds close to the SMBH with
near zero velocity. Therefore, in the bottom panel of
\fig\ref{fm_500_r_1_fm_6000_r_20}, we present a plot with a launch
radius of $r_1$ = 20 pc, a force multiplier of 6000, and a column
density of $N_{20}$ = 1. The maximum velocity reached is $\sim $
1700\kms with $n_{med}$ = 10. The best curve to represent the data
seems to be the one with $n_{med}$ = 0.33, whose maximum velocity is
$\sim $ 1500\kms, although the turnover point at $\sim $ 60 pc is
still too low according to our kinematic model. Furthermore, we have
had to tweak the launch radius and force multiplier to very specific
values, such that only a narrow range of the observed values gives a
reasonably decent fit. However, to directly test this velocity
profile, we generate biconical models similar to our previous
kinematic models (Paper II), to determine whether a more dynamical
velocity law, rather than the simple linear law, can reasonably match
the data.

\subsection{Dynamical Fit}
\begin{figure}[!ht]
  \vspace*{7mm}
  \centering
  \includegraphics[width=0.45\textwidth,height=0.29\textheight]
  {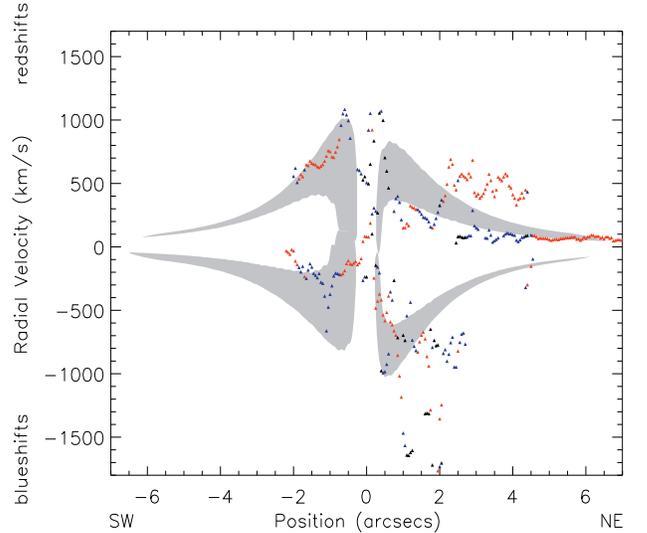}
  \caption{The model of slit 4 generated with input
    parameters from \tab\ref{tabsevenA}, and the best-fit velocity
    profile from the bottom panel of \fig\ref{fm_500_r_1_fm_6000_r_20}
    with $n_{med}$ = 0.33. Clearly this model is a poor match to the
    data.\vspace*{2mm}}
  \label{r20fm6000NH3model}
\end{figure}

We applied our kinematic models of Paper II with our dynamical
velocity law to the data. Previously we had used a simple kinematic
velocity law, which is based on the relation $v=kr$. Now instead we
substitute the velocity relation based on \eqn\ref{diff}, which
represented a more physical situation. A model with the new velocity
law of \eqn\ref{diff} is presented in \fig\ref{r20fm6000NH3model},
using the input parameters from \tab\ref{tabsevenA},
\begin{deluxetable}{cccccr}
  \tablenum{1}
  \tablewidth{0.45\textwidth}
  \tablecaption{\textsc{Parameters used to generate the drag model shown
      in \fig\ref{r20fm6000NH3model}.}
    \label{tabsevenA}}
  \tablehead{
    \colhead{$ z_{\mathrm{max}} $}  & \colhead{$
      \mathrm{\theta_{inner}} $}    &
    \colhead{$ \mathrm{\theta_{outer}} $} &
    \colhead{$ i_{\mathrm{axis}} $} & \colhead{$ PA_{\mathrm{axis}} $} &
    \colhead{$ v_{law}$}\\
    \colhead{(pc)}  & \colhead{(deg)} & \colhead{(deg)} &
    \colhead{(deg)} & \colhead{(deg)} & \colhead{}}
  \startdata
  450 & 10 & 40 & 5 & 57.8 & \eqn\ref{diff}\\
  \enddata
  \tablenotetext{\ }{\!\!\!\!$z_{max}$ is the distance from the center to
    one end of the bicone, along the z-axis.\\
    $\theta_{inner}$ is the inner opening angle of the bicone.\\
    $\theta_{outer}$ is the outer opening angle of the bicone.\\
    $i_{axis}$ is the inclination of the bicone axis out
    of the plane of the sky, with positive inclination implied by the
    north bicone closer toward the observer.\\
    $PA_{axis}$ is the position angle of the bicone axis in
    the plane of the sky.\\
    $v_{law}$ is the velocity law used in mapping the
    velocity field onto the bicone.}
\end{deluxetable}
and using the best-fit curve from the bottom panel of
\fig\ref{fm_500_r_1_fm_6000_r_20}, with $N_{20}$ = 1 and $n_{med}$ =
0.33. Our biconical model was constructed with an inner and outer half
opening angle of 10$\arcdeg$ and 40$\arcdeg$ respectively, and
inclination of 5$\arcdeg$, and position angle of 57.8$\arcdeg$, and a
half-size of 450 pc. Points on the bicone were assigned velocities
according to \eqn\ref{diff}. The center slit (slit 4, see Paper II) is
used here for comparison. We extracted the velocities from the model
at a position equivalent to slit 4 and those are shown in the shaded
regions in \fig\ref{r20fm6000NH3model}. The data from slit 4 are shown
in small triangles.

The model represents a poor fit to the data from slit 4 (the center
slit) of NGC~1068. This was expected from looking at the previous
velocity plots, as the turnover point was too low, the launch radius
was a bit large, and the velocity profile did not resemble our
kinematically derived linear profiles. We have varied the input
parameters from \tab\ref{tabsevenA}, but this makes very little
improvement. The turnover point, starting distance, and maximum
velocity of the bicone cannot be varied without changing the drag
parameters, as these are implicitly defined in \eqn\ref{diff}.  The
thickness of the bicone can be increased to accommodate more data
points, but the bicone will show lots of unnecessary shaded regions.
Changing the maximum extent of the bicone $z_{max}$ will have
absolutely no effect on the shaded region, except interrupting the
shaded regions before or continuing them beyond $\pm$ 6\arcsec. That
leaves us with only two parameters to vary, the inclination of the
bicone axis, and its position angle in the sky, and varying these two
alone did not fix the model.

\section{Conclusions}
With radiation pressure driving the NLR clouds, their velocities will
accelerate very quickly, within a few parsecs of the nucleus, assuming
the clouds are indeed launched close in. With the introduction of the
drag forces, the overall velocities are lowered, but even so the
velocities reach maximum too quickly. The data suggest that the clouds
are gradually increasing their velocities to a maximum at about 140
pc. This gradient in the velocity cannot be simply accounted for by
radiative forces driving the clouds. It seems therefore, that
radiation pressure may not be the only driving mechanism for the NLR
clouds, or that other forces are involved to steer the clouds to that
particular gradient. The velocity profile shown in the data resembles
one with a linear or `Hubble flow' law.  The inclusion of a drag force
only serves to reduce the overall high velocities due to radiation
driving, but the maximum velocities are reached very close to the
nucleus, too close to match the data effectively.

Gravitational forces alone cannot stop the fast moving clouds observed
in the NLR of NGC~1068. With velocities as high as 1500\kms at 140 pc,
the clouds should have escaped the NLR. To compound the problem, when
radiation forces are added, the cloud velocities are boosted even
more. Yet we see clouds that are slowing down and gradually reaching
systemic velocity. Therefore, the data suggest that there is a
powerful force dragging on the clouds to slow their velocity.

The drag force that we introduce can have a significant effect on the
clouds' velocities. We can conclude therefore that the drag forces are
a strong competitor to the radiative forces, strong enough to bring
the clouds to a halt even close to the nucleus, depending on the
column densities of the outflowing clouds and the densities of the
ambient medium. However the overall velocity profiles generated with
radiative, gravitational, and drag forces do not match the data for
NGC~1068. Assuming that the mass profile of NGC~4151 is similar to the
one for NGC~1068, it will prove difficult to match its observed
velocity profile, because the same linear trends are seen in the
velocity of the outflowing clouds. The same can be said for Mkn 3
\citep{Ruiz2001}.

\acknowledgments We would like to thank Douglas R. Gies and Paul J.
Wiita for providing insights into some of our modeling work. We would
also like to thank John. E. Everett for helpful discussions concerning
the driving mechanisms of the NLR gas. The data used in this paper
were obtained from the Multimission Archive at the Space Telescope
Science Institute (MAST).  STScI is operated by the Association of
Universities for Research in Astronomy, Inc., under NASA contract
NAS5-26555. This research has made use of NASA's Astrophysics Data
System.

\end{document}